\documentclass[a4paper,10.5pt]{article}
\usepackage{theorem}
\usepackage{latexsym,amssymb,amsfonts,amsmath,cite}
\usepackage[dvips]{graphicx}
\def\vector#1{\mbox{\boldmath $#1$}}

\newtheorem{definition}{Definition}
\newtheorem{theorem}{Theorem}
\newtheorem{lemma}{Lemma}
\newtheorem{corollary}[theorem]{Corollary}
\newtheorem{proposition}[theorem]{Proposition}

\begin{document}
\title{{\Large {\bf The stationary measure of a space-inhomogeneous quantum walk on the line 
}
}}

\author{ 
{\small 
Takako Endo$^{1}$ 
\footnote{g1170615@edu.cc.ocha.ac.jp
}\quad  
Norio Konno,$^{2}$ 
\footnote{konno@ynu.ac.jp 
}\quad 
}\\ 
{\scriptsize $^{1}$ 
Department of Physics, Graduate School of Humanities and Sciences, Ochanomizu University
}\\
{\scriptsize  Bunkyo, Tokyo 112-0012, Japan
} \\
{\scriptsize $^2$ 
Department of Applied Mathematics, Faculty of Engineering, Yokohama National University
}\\
{\scriptsize Hodogaya, Yokohama 240-8501, Japan
} \\
} 

\vskip 1cm

\date{\empty }
\pagestyle{plain}

\maketitle

\par\noindent
\begin{small}
\baselineskip=24pt
\par\noindent
{\bf Abstract}. 
We study a discrete-time quantum walk (QW) on the line with a single phase at the origin which was introduced and 
studied by  Wojcik et al.$[1]$. We call the model ``the Wojcik model" here.
Konno et al.$[2]$ investigated other types of QWs with one defect at the origin.
They presented a method which gives the stationary measure corresponding to localization for 
the QWs by use of the generating functions splitted in positive and negative parts respectively.
In this paper, we call the method ``the splitted generating function method (the SGF method)".
To clarify in detail which QW is appropriate for the SGF method might be one of the important challenges to investigate localization 
properties for various QWs.
As for the Wojcik model, we solve the eigenvalue problem by the SGF method and our results agree with Ref.$[1]$.
From the solution of the problem, we derive a stationary measure with an exponential decay for the position.
The explicit expression for the stationary measure is symmetric for the origin and 
ensures localization depending on the initial coin state.
\footnote[0]{
\par\noindent
$2010$ Mathematics Subject Classification: $60$B$10$, $81$P$45$, $81$P$68$.
}

\end{small}

\baselineskip=24pt

\setcounter{equation}{0}
\section{Introduction}

Quantum walks (QWs) can be regarded as quantum analogue of the classical random walks. 
Very recently QWs have been widely investigated by a number of groups in connection with various topics, 
for example, the quantum computing $[3,4]$, physical systems $[5,6]$.
There are some reviews on QWs, such as $[7,8]$.
Up to now, the asymptotic behavior of QWs has been intensively studied $[2,9,10,11,12,13,14]$.
There are two kinds of limit theorems to understand the asymptotic behavior. 
One is the time-averaged limit measure corresponding to localization. 
As Konno et al.$[2]$ discussed, the measure is closely related to the stationary measure for some one-dimensional two-state QWs with one defect.
The other is the rescaled weak limit theorem corresponding to the ballistic spreading $[2]$.  
There are four analytical methods in main to investigate the asymptotic behavior of QWs, Fourier's method $[11]$, the stationary phase method $[12]$, 
the CGMV method $[13]$, and so-called the generating function method $[14]$. 
In this paper, we use other method in which 
the generating function for the probability amplitude, splitted in positive and negative parts respectively, plays an important role. 
We call the method ``the splitted generating function method (the SGF method)" which was introduced in Konno et al.$[2]$. 
The method is useful to obtain the stationary measure for 
the QW with one defect on the line. 
However, answers for the following questions are not known: what types of QW models are appropriate for the method and how the initial coin states influence the stationary measure.
Wojcik et al.$[1]$ reported that giving the phase at a single point in QW exhibits an astonishing localization effect.
In this paper, we call the model ``Wojcik model".
We solve the eigenvalue problem of the Wojcik model taking advantage of the SGF method.
The solution of the problem leads to a stationary measure.
The expression for the stationary measure depending on the initial state is symmetric for the origin.
In addition, the measure has an exponential decay for the position.
We mention that to compute the time-averaged limit measure of the Wojcik model and clarify the relation between the limit and stationary measures is one of 
the interesting future problems.

Before we discuss the Wojcik model, we introduce a discrete time space-inhomogeneous QW on the line 
which is a quantum version of the classical random walk with an additional coin state. 
The Wojcik model can be regarded as a special case for the QW.
The particle has a coin state at time $n$ and position $x$ described by a two dimensional vector: 
\[\Psi_{n}(x)=
\begin{bmatrix}
\Psi^{L}_{n}(x)\\
\Psi^{R}_{n}(x)
\end{bmatrix}\quad(x\in\mathbb{Z}),\]
where $\mathbb{Z}$ is the set of the integers.
The upper and lower elements express left and right chiralities respectively.
The time evolution is determined by $2\times 2$ unitary matrices $U_{x}$ :

\[U_{x}=\begin{bmatrix}
a_{x}&b_{x}\\
c_{x}&d_{x}\\
\end{bmatrix}\;\;\;(x\in \mathbb{Z}).\]\\
The subscript $x$ stands for the location.
We devide $U_{x}$ into $U_{x}=P_{x}+Q_{x}$,
where
\[ 
P_{x}=\begin{bmatrix} 
a_{x} & b_{x} \\
0 & 0 
\end{bmatrix}, \quad
Q_{x}=\begin{bmatrix}
0 & 0 \\
c_{x} & d_{x}
\end{bmatrix}.\]
The $2\times2$ matrix $P_{x}$ (resp. $Q_{x}$) represents that the walker moves to the left (resp. right) at position $x$ 
at each time step.
Then the time evolution of the walk is defined by 
\[\Psi_{n+1}(x)= P_{x+1}\Psi_{n}(x+1)+ Q_{x-1}\Psi_{n}(x-1).\]
That is \[\begin{bmatrix}
\Psi_{n+1}^{L}(x)\\
\Psi_{n+1}^{R}(x)\\
\end{bmatrix}=\begin{bmatrix}           
a_{x+1}\Psi_{n}^{L}(x+1)+b_{x+1}\Psi_{n}^{R}(x+1)\\
c_{x-1}\Psi_{n}^{L}(x-1)+d_{x-1}\Psi_{n}^{R}(x-1)

\end{bmatrix}.\]
Now let
\[\Psi_{n}= {}^T\![\cdots,\Psi_{n}^{L}(-1),\Psi_{n}^{R}(-1),\Psi_{n}^{L}(0),\Psi_{n}^{R}(0),\Psi_{n}^{L}(1),\Psi_{n}^{R}(1),\cdots ],\]
and
\[U^{(s)}=\begin{bmatrix}
\ddots&\vdots&\vdots&\vdots&\vdots&\cdots\\
\cdots&O&P_{-1}&O&O&O\cdots\\
\cdots&Q_{-2}&O&P_{0}&O&O\cdots\\
\cdots&O&Q_{-1}&O&P_{1}&O\cdots\\
\cdots&O&O&Q_{0}&O&P_{2}\cdots\\
\cdots&O&O&O&Q_{1}&O\cdots\\
\cdots&\vdots&\vdots&\vdots&\vdots&\ddots
\end{bmatrix}\;\;\;
with\;\;\;O=\begin{bmatrix}0&0\\0&0\end{bmatrix},\]
where $T$ means the transposed operation. Then the state of the QW at time $n$ is given by
$\Psi_{n}=(U^{(s)})^{n}\Psi_{0}$ for any $n\geq0$. 
Let $\mathbb{R}_{+}=[0,\infty)$. Here we introduce a map 
$\phi:(\mathbb{C}^{2})^{\mathbb{Z}}\rightarrow \mathbb{R}_{+}^{\mathbb{Z}}$
such that for
\[\Psi= {}^T\!\left[\cdots,\begin{bmatrix}
\Psi^{L}(-1)\\
\Psi^{R}(-1)\end{bmatrix},\begin{bmatrix}
\Psi^{L}(0)\\
\Psi^{R}(0)\end{bmatrix},\begin{bmatrix}
\Psi^{L}(1)\\
\Psi^{R}(1)\end{bmatrix},\cdots\right]\in(\mathbb{C}^{2})^{\mathbb{Z}},\]
we define the measure of the QW by
\[\mu:\mathbb{Z}\to\mathbb{R}_{+}\;such\;that\;\mu(x)=\phi(\Psi(x)) = |\Psi^{L}(x)|^{2} + |\Psi^{R}(x)|^{2}\;\;\;(x\in\mathbb{Z}).\]
We should note that $\mu(x)$ gives the measure of the QW at position $x$.
Now we are ready to introduce the stationary measure.
Put 
\begin{align}
\Sigma_{s}=\{\phi(\Psi_{0})\in\mathbb{R}_{+}^{\mathbb{Z}}: there\;exists\;\Psi_{0}
\;such\;that\;\;\phi((U^{(s)})^{n}\Psi_{0})=\phi(\Psi_{0})\;for\;any\;n\geq 0\},\end{align}
and we call the element of $\Sigma_{s}$ the stationary measure of the QW.

The rest of this paper is organized as follows. In Sect. $2$, we introduce the SGF method and the Wojcik model.
In Sect. $3$, we solve the eigenvalue problem (Proposition $1$) and present a stationary measure (Theorem $2$).
Sect. $4$ deals with the proof of Proposition $1$.

\section{Method and model}

First of all, let us introduce the SGF method briefly.
In general, the time evolution of a space-inhomogeneous QW on the line is determined by its unitary matrix:
\begin{align}
U_{x}=\begin{bmatrix}
a_{x}&b_{x}\\
c_{x}&d_{x}
\end{bmatrix}
\quad(x\in\mathbb{Z},\;a_{x},b_{x},c_{x},d_{x}\in\mathbb{C}),
\end{align} 
and its initial state:  
\begin{align}\Psi_{0}= {}^T\![\cdots,\Psi_{0}^{L}(-1),\Psi_{0}^{R}(-1),\Psi_{0}^{L}(0),\Psi_{0}^{R}(0),\Psi_{0}^{L}(1),\Psi_{0}^{R}(1),\cdots ].\end{align}
We should note that the coin state at time $n$ and position $x$ is given by
\[\Psi_{n}(x)=
\begin{bmatrix}
\Psi^{L}_{n}(x)\\
\Psi^{R}_{n}(x)
\end{bmatrix}\in\mathbb{C}^{2},\]
which is also called ``probability amplitude".
Here we devide $U_{x}$ into $2\times2$ matrices $P_{x}$ and $Q_{x}$ as
\[U_{x}=P_{x}+Q_{x},\]
where
\[P_{x}=\begin{bmatrix}a_{x}&b_{x}\\ 0&0\end{bmatrix},\quad Q_{x}=\begin{bmatrix}0&0\\c_{x}&d_{x}\end{bmatrix}.\]
Recall that matrix $P_{x}$ (resp. $Q_{x}$) represents that the walker moves to the left (resp. right) at position $x$ at each time step.
Let us consider the eigenvalue problem:
\begin{align}
U^{(s)}\Psi=\lambda\Psi,
\end{align}
where $\lambda\in\mathbb{C}$ with $|\lambda|=1$ and $U^{(s)}$ is an $\infty\times\infty$ unitary matrix. Here $U^{(s)}$ defines the time evolution of the model. 
If we assume that the initial state $\Psi_{0}$ is a solution of Eq.$(2.4)$, we have
\[\Psi_{n}=(U^{(s)})^{n}\Psi_{0}=\lambda^{n}\Psi_{0}.\]
Noting that $|\lambda|=1$, we see
\begin{align*}
\mu_{n}(x)=\|\Psi_{n}(x)\|^{2}
=|\lambda|^{2n}\|\Psi_{0}(x)\|^{2}=\mu_{0}(x)\quad
(x\in\mathbb{Z}).
\end{align*}
Therefore $\mu_{0}(x)=\phi(\Psi_{0}(x))$ gives the stationary measure.

In this paper, localization for a discrete time QW on the line is defined as follows.

\begin{definition}
Localization for the QW starting from the origin happens when
\[\limsup_{n\to\infty}\mu_{n}(0)>0.\]
\end{definition}

By the SGF method, the stationary measure for the QW is given in the following way. 
To begin with, we introduce the generating functions of $\Psi^{L}(x)$ and $\Psi^{R}(x)$ which are the key elements for 
our discussion:
\[f^{j}_{+}(z)=\sum_{x=1}^{\infty}\Psi^{j}(x)z^{x},\quad f^{j}_{-}(z)=\sum_{x=-1}^{-\infty}\Psi^{j}(x)z^{x}\quad(j=L,R).\]

We consider the solution of Eq. $(2.4)$ $\Psi(x)={}^T\![\Psi^{L}(x),\Psi^{R}(x)]$ satisfies
\begin{align}
\lambda\begin{bmatrix}\Psi^{L}(x)\\ \Psi^{R}(x)\end{bmatrix}
=\begin{bmatrix}0&0\\c_{x-1}&d_{x-1}\end{bmatrix}\begin{bmatrix} \Psi^{L}(x-1)\\ \Psi^{R}(x-1)\end{bmatrix}+
\begin{bmatrix}a_{x+1}&b_{x+1}\\0&0\end{bmatrix}\begin{bmatrix} \Psi^{L}(x+1)\\ \Psi^{R}(x+1)\end{bmatrix}.
\end{align}
We should note that Eq.$(2.5)$ is equvalent to Eq.$(2.4)$.
Then we obatin the probability amplitude $\Psi(x)$ by use of the generating functions.

From now on, we focus on the Wojcik model, whose time evolution is determined by
\begin{align}
U_{x}=\left\{ \begin{array}{ll}
\dfrac{1}{\sqrt{2}}\begin{bmatrix} 1&1\\ 1& -1\end{bmatrix}& (x=\pm1,\pm2,\cdots), \\\
\dfrac{\omega}{\sqrt{2}}\begin{bmatrix} 1&1\\ 1& -1\end{bmatrix}& (x=0), \\
\end{array} \right.
\end{align}
with $\omega=e^{2i\pi\phi}$ where $\phi\in(0,1)$.
The Hadamard walk, given by $\phi\to0$ in Eq.$(2.6)$, has been attracted much attention for a decade $[2,7,8,9]$.
We mention that our model has a phase $2\pi\phi$ only at the origin.
Remark that Wojcik et al.$[1]$ solved the eigenvalue problem $(U^{(s)})^{2}\Psi=\lambda^{2}\Psi$ by using recurrence equations. 

\section{Results}

Applying the SGF method to the Wojcik model, we solve the eigenvalue problem $U^{(s)}\Psi=\lambda\Psi$ as follows.
\newpage
\begin{proposition}
Let $\Psi(x)={}^T\![\Psi^{L}(x),\Psi^{R}(x)]$ be the probability amplitude. 
Put $\alpha=\Psi^{L}(0)$ and $\beta=\Psi^{R}(0)$.
Then the solutions for 
\[U^{(s)}\Psi=\lambda\Psi,\]
where 
$\lambda\in\mathbb{C}$ with $|\lambda|=1$,
are
\begin{eqnarray}
\Psi(x)=(-\theta_{s}\operatorname{sgn}(x))^{|x|}\times
\left\{\begin{array}{ll}
\begin{bmatrix}\alpha\\(1-\omega)\alpha+\omega\beta\end{bmatrix} & (x\geq 1), \\\
\begin{bmatrix}\alpha\\ \beta\end{bmatrix} & (x=0), \\\
\begin{bmatrix}(\omega-1)\beta+\omega\alpha \\ \beta\end{bmatrix} & (x\leq -1), 
\end{array}\right.
\end{eqnarray} 
with $\beta^{2}=-\alpha^{2}$, that is, $\beta=i \alpha$ or $\beta=-i \alpha$. Here, $\omega=e^{2\pi i\phi}\;(\phi\in(0,1))$. 
\par\noindent
(1) $\beta=i \alpha$ case.
\begin{align}
\lambda^{2}&=\dfrac{\omega(1-2\omega+\omega^{2})-i\omega(1-\omega+\omega^{2})}{1-2\omega+2\omega^{2}},
\nonumber
\\
\theta_{s}^{2}&=\dfrac{\omega}{\omega^{2}-3\omega+1-i(\omega^{2}-1)} =\dfrac{1}{2\cos(2\pi\phi)+2\sin(2\pi\phi)-3}.
\end{align}
(2) $\beta= -i \alpha$ case.
\begin{align}
\lambda^{2} 
&=\dfrac{\omega(1-2\omega+\omega^{2})+i\omega(1-\omega+\omega^{2})}{1-2\omega+2\omega^{2}},
\nonumber
\\
\theta_{s}^{2} 
&=\dfrac{\omega}{\omega^{2}-3\omega+1+i(\omega^{2}-1)} =\dfrac{1}{2\cos(2\pi\phi)-2\sin(2\pi\phi)-3}.
\end{align}
\end{proposition}
The proof of Proposition $1$ is given in Sect. $4$. 
We see that Eqs.$(3.8)$ and $(3.9)$ agree with $x_{-}$ and $x_{+}$ in Eq.$(11)$ of Ref.$[1]$ respectively. \\
Noting that $|\alpha|=|\beta|$ and
\[\mu(x)=\|\Psi(x)\|^{2}=|\Psi^{L}(x)|^{2}+|\Psi^{R}(x)|^{2},\]
we obtain the stationary measure for the Wojcik model as follows.
\begin{theorem} 
We have the stationary measure as
\[\mu(x)=\|\Psi(x)\|^{2}=2|\alpha|^{2}|\theta_{s}|^{2|x|}
\times\left\{\begin{array}{ll}\Gamma(\phi)&(x\neq0),\\
1&(x=0),\end{array}\right.\]
where
\begin{align*}
\Gamma(\phi)=\left\{ \begin{array}{ll}
2-\cos(2\pi\phi)-\sin(2\pi\phi)&(\beta=i\alpha),\\
2-\cos(2\pi\phi)+\sin(2\pi\phi)&(\beta=-i\alpha),
\end{array} \right.
\end{align*}
and
\begin{align}|\theta_{s}|^{2}=\left\{ \begin{array}{ll}
\dfrac{1}{3-2\cos(2\pi\phi)-2\sin(2\pi\phi)} & (\beta=i\alpha), \\
\dfrac{1}{3-2\cos(2\pi\phi)+2\sin(2\pi\phi)} & (\beta=-i\alpha). 
\end{array} \right.\end{align}
\end{theorem}
\par\indent
We emphasize that the stationary measure is symmetric with respect to the origin.
The result also suggests that the localization effect depends on the choice of the parameter $\phi$ 
and the probability amplitude $\Psi(0)={}^T\![\alpha,\beta]$.
We notice that localization happens except for $\alpha=\beta=0$.
We also emphasize that the stationary measure has an exponential decay for the position.
Our results imply that the SGF method is effective for the Wojcik model.

Next we show the relation between the parameters $\phi$ and $\xi$.
We should recall that $\omega=e^{2\pi i\phi}\;(\phi\in(0,1))$ and $\lambda=e^{i\xi}\;(\xi\in\mathbb{R})$.
Here we put $\omega=C+iS$, that is, \[C=\cos(2\pi\phi)\quad and\quad S=\sin(2\pi\phi).\]
We will express $\lambda^{2}=e^{2i\xi}$ in terms of the parameter $\phi$ by using Proposition $1$.
Here $\lambda$ is a solution for the eigenvalue problem
\[U^{(s)}\Psi=\lambda\Psi.\]
Note that $\alpha=\Psi^{L}(0)$ and $\beta=\Psi^{R}(0)$.
\begin{corollary}
\begin{enumerate}
\item $\beta=i\alpha$ case. We have
\begin{align*}\left\{ \begin{array}{l}
\cos(2\xi)=\dfrac{-2+6C+6S-6CS-8C^{2}+4C^{3}-4S^{3}}{5-12C+8C^{2}}, \\\
\sin(2\xi)=\dfrac{1-4C+8S-8CS+6C^{2}-4C^{3}-4S^{3}}{5-12C+8C^{2}}. 
\end{array} \right.\end{align*}
\item $\beta=-i\alpha$ case. We have
\begin{align*}\left\{ \begin{array}{l}
\cos(2\xi)=\dfrac{-2+6C-6S+6CS-8C^{2}+4C^{3}+4S^{3}}{5-12C+8C^{2}}, \\\
\sin(2\xi)=\dfrac{-1+4C+8S-8CS-6C^{2}+4C^{3}-4S^{3}}{5-12C+8C^{2}}. 
\end{array} \right.\end{align*}
\end{enumerate}
\end{corollary}
\noindent
\par\indent
Here we show some examples to confirm that Corollary $3$ for $\beta=i\alpha$ case is correct.
\begin{enumerate}
\item $\phi\to0$ ($\omega\to 1$) case.\\
We see $C\to1$ and $S\to0$, and obtain
$\cos(2\xi)\to0$ and $\sin(2\xi)\to-1$. \\
Hence we have $\lambda^{2}\to-i$, and the result agrees with that of Eq.$(8)$ in Ref.$[1]$.
\item $\phi=\dfrac{1}{4}$ ($\omega=i$) case.\\
We see $C=0$ and $S=1$ and we have $\cos(2\xi)=0$ and $\sin(2\xi)=1$.\\ 
Therefore we have $\lambda^{2}=i$, and the result agrees with that of Eq.$(8)$ in Ref.$[1]$.

\end{enumerate}

\section{Proof of Proposition $1$}
In this section, we prove of Proposition $1$. 
Let us start with the eigenvalue problem:
\begin{align}U^{(s)}\Psi=\lambda\Psi,\end{align}
where $\lambda\in\mathbb{C}$ with $|\lambda|=1$.
We should recall that 
\[U^{(s)}=\begin{bmatrix}
\ddots&\vdots&\vdots&\vdots&\vdots&\cdots\\
\cdots&O&P_{-1}&O&O&O\cdots\\
\cdots&Q_{-2}&O&P_{0}&O&O\cdots\\
\cdots&O&Q_{-1}&O&P_{1}&O\cdots\\
\cdots&O&O&Q_{0}&O&P_{2}\cdots\\
\cdots&O&O&O&Q_{1}&O\cdots\\
\cdots&\vdots&\vdots&\vdots&\vdots&\ddots
\end{bmatrix}\;\;\;
with\;\;\;O=\begin{bmatrix}0&0\\0&0\end{bmatrix}.\]
We solve the eigenvalue problem $(4.11)$ by the SGF method.
As another expression for Eq.$(4.11)$, we have
\begin{align}
\lambda\Psi(x)=P_{x+1}\Psi(x+1)+Q_{x-1}\Psi(x-1),
\end{align}
where 
\begin{align*}
P_{x}=\left\{ \begin{array}{ll}
\dfrac{1}{\sqrt{2}}\begin{bmatrix}1&1\\ 0&0 \end{bmatrix} &(x=\pm1,\pm2,\cdots), \\\
\dfrac{\omega}{\sqrt{2}}\begin{bmatrix}1&1\\ 0&0 \end{bmatrix}& (x=0),
\end{array} \right.and\;\;
Q_{x}=\left\{\begin{array}{ll}
\dfrac{1}{\sqrt{2}}\begin{bmatrix}0&0\\ 1&-1 \end{bmatrix} &(x=\pm1,\pm2,\cdots), \\\
\dfrac{\omega}{\sqrt{2}}\begin{bmatrix}0&0\\ 1&-1 \end{bmatrix}& (x=0).
\end{array} \right.
\end{align*}
Note that $\omega=e^{2\pi i\phi}$ with $\phi\in(0,1)$.

Let $\Psi(x)={}^T\![\Psi^{L}(x),\;\Psi^{R}(x)]\;(x\in\mathbb{Z})$
be the probability amplitude of the model. 
Then we construct the equation of the evolution in terms of $\lambda, \Psi^{L}(x)$, and $\Psi^{R}(x)$:
\begin{enumerate}
\item $x\neq\pm1$ case. 
\begin{align}\left\{ \begin{array}{l}
\lambda\Psi^{L}(x)=\dfrac{1}{\sqrt{2}}\Psi^{L}(x+1)+\dfrac{1}{\sqrt{2}}\Psi^{R}(x+1),\\
\lambda\Psi^{R}(x)=\dfrac{1}{\sqrt{2}}\Psi^{L}(x-1)-\dfrac{1}{\sqrt{2}}\Psi^{R}(x-1).\\
\end{array} \right. 
\end{align}
\item $x=1$ case. 
\begin{align}\left\{ \begin{array}{l}
\lambda\Psi^{L}(1)=\dfrac{1}{\sqrt{2}}\Psi^{L}(2)+\dfrac{1}{\sqrt{2}}\Psi^{R}(2),\\
\lambda\Psi^{R}(1)=\dfrac{\omega}{\sqrt{2}}\Psi^{L}(0)-\dfrac{\omega}{\sqrt{2}}\Psi^{R}(0).\\
\end{array} \right. 
\end{align}
\item $x=-1$ case. 
\begin{align}\left\{ \begin{array}{l}
\lambda\Psi^{L}(-1)=\dfrac{\omega}{\sqrt{2}}\Psi^{L}(0)+\dfrac{\omega}{\sqrt{2}}\Psi^{R}(0),\\
\lambda\Psi^{R}(-1)=\dfrac{1}{\sqrt{2}}\Psi^{L}(-2)-\dfrac{1}{\sqrt{2}}\Psi^{R}(-2).\\
\end{array} \right. 
\end{align}
\end{enumerate}

Here we introduce the generating functions for $\Psi^{L}(x)$ and $\Psi^{R}(x)$:
\begin{align}
f^{j}_{+}(z)=\sum_{x=1}^{\infty} \Psi^{j}(x)z^{x},\quad f^{j}_{-}(z)=\sum_{x=-1}^{-\infty} \Psi^{j}(x)z^{x}\quad(j=L,R).
\end{align}
Then combining the generating functions with Eqs.$(4.13),(4.14)$ and $(4.15)$, we obtain the following lemma.
The proof of Lemma $1$ is given in Appendix.
\begin{lemma}
Put
\begin{align*}A=\begin{bmatrix}\lambda-\dfrac{1}{\sqrt{2}z}& -\dfrac{1}{\sqrt{2}z}\\
-\dfrac{z}{\sqrt{2}}& \lambda+\dfrac{z}{\sqrt{2}}
\end{bmatrix},\;\vector{f}_{\pm}(z)=\begin{bmatrix}f^{L}_{\pm}(z)\\ f_{\pm}^{R}(z)\end{bmatrix},\\
\vector{a}_{+}(z)=\begin{bmatrix}-\lambda\alpha\\ \dfrac{\omega z(\alpha-\beta)}{\sqrt{2}}\end{bmatrix},
\quad \vector{a}_{-}(z)=\begin{bmatrix}\dfrac{\omega(\alpha+\beta) }{\sqrt{2}z}\\-\lambda\beta\end{bmatrix},\end{align*}
where $\omega=e^{2\pi i\phi}\;(\phi\in(0,1)),\;\alpha=\Psi^{L}(0)$, and $\beta=\Psi^{R}(0)$.
Then we have
\begin{align}
A\vector{f}_{\pm}(z)=\vector{a}_{\pm}(z).\;
\end{align}
\end{lemma}

Noting that
\begin{align}\det A=\dfrac{\lambda}{\sqrt{2}z}\left\{z^{2}-\sqrt{2}\left(\dfrac{1}{\lambda}-\lambda\right)z-1\right\},\end{align}
we put $\theta_{s}$ and $\theta_{l}\in\mathbb{C}$ which satisfy
\begin{align}
\det A=\dfrac{\lambda}{\sqrt{2}z}(z-\theta_{s})(z-\theta_{l}),
\end{align}
and $|\theta_{s}|\leq1\leq|\theta_{l}|$. Combining Eq.$(4.18)$ with Eq.$(4.19)$ gives $\theta_{s}\theta_{l}=-1$.

Now let us derive $f_{\pm}^{L}(z)$ and $f_{\pm}^{R}(z)$ from Lemma $1$.
\begin{enumerate}
\item Consider $f_{+}^{L}(z)$. Eq.$(4.17)$ implies
\begin{eqnarray*}
f^{L}_{+}(z)\!\!\!&=&\!\!\!\dfrac{1}{\det A}\left\{\left(\lambda+\dfrac{z}{\sqrt{2}}\right)(-\lambda\alpha)+\dfrac{\omega}{2}(\alpha-\beta)\right\}\\
\!\!\!&=&\!\!\!\dfrac{1}{\det A}\left(-\dfrac{\lambda \alpha}{\sqrt{2}}\right)\left[z-\dfrac{\sqrt{2}}{\lambda\alpha}\left\{-\lambda^{2}\alpha+\dfrac{\omega}{2}(\alpha-\beta)\right\}\right].
\end{eqnarray*}
If we put 
$\theta_{s}=\dfrac{\sqrt{2}}{\lambda\alpha}\left\{\left(-\lambda^{2}+\dfrac{\omega}{2}\right)\alpha-\dfrac{\omega}{2}\beta\right\}$, we have
\begin{eqnarray*}
f^{L}_{+}(z)\!\!\!&=&\!\!\!-\dfrac{\alpha z}{z-\theta_{l}}
=-\dfrac{\alpha z}{z+\dfrac{1}{\theta_{s}}}
=-\alpha\dfrac{z\theta_{s}}{z\theta_{s}+1}\\
\!\!\!&=&\!\!\!-\alpha(\theta_{s}z)\{1+(-\theta_{s}z)+(-\theta_{s}z)^{2}+(-\theta_{s}z)^{3}+\cdots\}.
\end{eqnarray*}
Hence we realize
\begin{align}
f^{+}_{L}(z)=\alpha\sum_{x=1}^{\infty}(-\theta_{s}z)^{x}.
\end{align}
Combining Eq.$(4.20)$ with the definition of $f_{+}^{L}(z)$, we obtain
\[
\Psi^{L}(x)=\alpha(-\theta_{s})^{x}\;\;\;(x=1,2,\cdots),
\]
with
\begin{align}
\theta_{s}=\dfrac{\sqrt{2}}{\lambda\alpha}\left\{\left(-\lambda^{2}+\dfrac{\omega}{2}\right)\alpha-\dfrac{\omega}{2}\beta\right\}.
\end{align}
\item Consider $f_{+}^{R}(z)$. Eq.$(4.17)$ gives 
\[
f^{R}_{+}(z)=\dfrac{1}{\det A}\dfrac{\lambda}{\sqrt{2}}\{(\omega-1)\alpha-\omega\beta\}\left\{z-\dfrac{\sqrt{2}}{\lambda}\dfrac{\dfrac{\omega}{2}(\alpha-\beta)}{(\omega-1)\alpha-\omega\beta}\right\}.
\]
If we put $\theta_{s}=\dfrac{\omega(\alpha-\beta)}{\sqrt{2}\lambda\{(\omega-1)\alpha-\omega\beta\}}$, we have
\begin{eqnarray}
f^{R}_{+}(z)=-\dfrac{\{(1-\omega)\alpha+\omega\beta\}z}{z-\theta_{l}}=
\{(1-\omega)\alpha+\omega\beta\}\sum_{x=1}^{\infty}(-\theta_{s}z)^{x}.
\end{eqnarray}
Combining Eq.$(4.22)$ with the definition of $f_{+}^{R}(z)$, we obtain
\[
\Psi^{R}(x)=\{(1-\omega)\alpha+\omega\beta\}(-\theta_{s})^{x}\;\;\;(x=1,2,\cdots),
\]
with
\begin{align}\theta_{s}=\dfrac{\omega(\alpha-\beta)}{\sqrt{2}\lambda\{(\omega-1)\alpha-\omega\beta\}}.\end{align}
\item Consider $f_{-}^{L}(z)$. Eq.$(4.17)$ leads to
\[
f^{L}_{-}(z)=\dfrac{\omega(\alpha+\beta)}{\sqrt{2}\lambda(z-\theta_{l})(z-\theta_{s})}\left[z-\dfrac{\sqrt{2}\lambda\{-\omega\alpha+(1-\omega)\beta\}}{\omega(\alpha+\beta)}\right].
\]
If we put $\theta_{l}=-\dfrac{\sqrt{2}\lambda\{\omega\alpha+(\omega-1)\beta\}}{\omega(\alpha+\beta)}$, we have
\begin{eqnarray*}
f^{L}_{-}(z)\!\!\!&=&\!\!\!\dfrac{\omega(\alpha+\beta)}{\sqrt{2}\lambda}\times\dfrac{1}{z-\theta_{s}}
=\dfrac{\omega(\alpha+\beta)}{\sqrt{2}\lambda}\times\dfrac{1}{\theta_{s}}\times\dfrac{\theta_{s}}{z}\dfrac{1}{1-\dfrac{\theta_{s}}{z}}\\
\!\!\!&=&\!\!\!\dfrac{\omega(\alpha+\beta)}{\sqrt{2}\lambda}\times\dfrac{1}{\theta_{s}}\left\{\dfrac{\theta_{s}}{z}+\left(\dfrac{\theta_{s}}{z}\right)^{2}+\cdots\right\}.
\end{eqnarray*}
Therefore we see
\begin{align}
f^{L}_{-}(z)=\dfrac{\omega(\alpha+\beta)}{\sqrt{2}\lambda}\times\left(\dfrac{1}{\theta_{s}}\right)\sum^{-\infty}_{x=-1}(\theta^{-1}_{s}z)^{x}
=\{\omega\alpha+(\omega-1)\beta\}\sum^{-\infty}_{x=-1}(\theta^{-1}_{s}z)^{x}.
\end{align}
Combining Eq.$(4.24)$ with the definition of $f^{L}_{-}(z)$, we obtain
\[
\Psi^{L}(x)=\{\omega\alpha+(\omega-1)\beta\}(\theta_{s})^{-x}\;\;\;(x=-1,-2,\cdots),
\]
with
\begin{align}
\theta_{s}=\dfrac{\omega(\alpha+\beta)}{\sqrt{2}\lambda\{\omega\alpha+(\omega-1)\beta\}}.
\end{align}
\item Consider $f_{-}^{R}(z)$. Eq.$(4.17)$ gives
\[
f^{R}_{-}(z)=\dfrac{\sqrt{2}z}{\lambda}\dfrac{1}{(z-\theta_{l})(z-\theta_{s})}\times\dfrac{\omega\alpha+(\omega-2\lambda^{2})\beta}{2z}\left(z-\dfrac{2\lambda\beta}{-\sqrt{2}\omega\alpha+\sqrt{2}(2\lambda^{2}-\omega)\beta}\right).
\]
If we put $\theta_{l}=\dfrac{2\lambda\beta}{-\sqrt{2}\omega\alpha+\sqrt{2}(2\lambda^{2}-\omega)\beta}$, we have
\begin{eqnarray*}
f^{R}_{-}(z)\!\!\!&=&\!\!\!\dfrac{\sqrt{2}}{\lambda}\left\{\dfrac{\omega}{2}\alpha+\left(\dfrac{\omega}{2}-\lambda^{2}\right)\beta\right\}\times\dfrac{1}{z-\theta_{s}}\\
\!\!\!&=&\!\!\!\dfrac{\sqrt{2}}{\lambda}\left\{\dfrac{\omega}{2}\alpha+\left(\dfrac{\omega}{2}-\lambda^{2}\right)\beta\right\}\times\dfrac{1}{\theta_{s}}\times\dfrac{\theta_{s}}{z}\times\dfrac{1}{1-\dfrac{\theta_{s}}{z}}.
\end{eqnarray*}
Hence we see
\begin{align}
f^{R}_{-}(z)=\beta\sum^{-\infty}_{x=-1}(\theta_{s}^{-1}z)^{x}.
\end{align}
Combining Eq.$(4.26)$ with the definition of $f^{R}_{-}(z)$, we obtain
\[
\Psi^{R}(x)=\beta(\theta_{s})^{-x}\;\;\;(x=-1,-2,\cdots),
\]
with
\begin{align}
\theta_{s}=\dfrac{\sqrt{2}}{\lambda\beta}\left\{\dfrac{\omega}{2}\alpha+\left(\dfrac{\omega}{2}-\lambda^{2}\right)\beta\right\}.
\end{align}
\end{enumerate} 
Summarizing the above discussions, we get
\begin{align}
\Psi(x)=\left\{ \begin{array}{ll}
(-\theta_{s})^{x}\begin{bmatrix}\alpha\\ (1-\omega)\alpha+\omega\beta\end{bmatrix} &(x=1,2,\cdots),\\
\begin{bmatrix}\alpha \\ \beta \end{bmatrix} &(x=0),\\
(\theta_{s})^{|x|}\begin{bmatrix}(\omega-1)\beta+\omega\alpha\\ \beta\end{bmatrix} &(x=-1,-2,\cdots).
\end{array} \right.
\end{align}
\par\indent
Here noting that Eqs.$(4.21)$, $(4.23)$, $(4.25)$ and $(4.27)$, which are the different forms of 
$\theta_{s}$, are equivalent, we see that there are only two cases: $\beta=i\alpha$ and $\beta=-i\alpha$.
The first case corresponds to Eq.$(12)$ in Ref.$[1]$: $\overline{\alpha_{0}^{(-)}}=C,\;\overline{\beta_{0}}^{(-)}=iC$.
On the other hand, the second case corresponds to Eq.$(12)$ in Ref.$[1]$:
$\overline{\alpha_{0}^{(+)}}=C,\;\overline{\beta_{0}}^{(+)}=-iC$.

From now on, we will express $\lambda^{2}$ and $\theta_{s}^{2}$ in terms of $\omega$ in each case.
From Eqs.$(4.21)$ and $(4.23)$, we obtain
\begin{align}
(-\lambda^{2}\omega+\lambda^{2}-\omega)\alpha+(-\omega^{2}+\omega+\lambda^{2}\omega)\beta=0.
\end{align} 
Rewriting Eq.$(4.29)$ for $\lambda$, we have
\begin{align}
(\alpha-\alpha\omega+\beta\omega)\lambda^{2}-\alpha\omega+\beta\omega(1-\omega)=0.
\end{align}
\begin{enumerate}
\item $\beta=i\alpha$ case.
Eq.$(4.30)$ implies
\begin{align}
\lambda^{2}=\dfrac{\omega(1-2\omega+\omega^{2})-i\omega(1-\omega+\omega^{2})}{1-2\omega+2\omega^{2}}.
\end{align}
Eq.$(4.31)$ agrees with $\lambda_{-}$ in Eq.$(8)$ of Ref.$[1]$.
\item $\beta=-i\alpha$ case.
Eq.$(4.30)$ gives
\begin{align}
\lambda^{2}=\dfrac{\omega(1-2\omega+\omega^{2})+i\omega(1-\omega+\omega^{2})}{1-2\omega+2\omega^{2}}.
\end{align}
Eq.$(4.32)$ also agrees with $\lambda_{+}$ in Eq.$(8)$ of Ref.$[1]$.
\end{enumerate}
Next, by Eq.$(4.25)$, we see
\begin{align}
\theta_{s}^{2}=\dfrac{\omega^{2}(\alpha+\beta)^{2}}{2\lambda^{2}\{\omega\alpha+(\omega-1)\beta\}^{2}}.
\end{align}
This equation implies that $\theta_{s}^{2}$ can be expressed by only $\omega$ as follows:
\begin{enumerate}
\item $\beta=i\alpha$ case.\\
\begin{align}
\theta_{s}^{2}=\dfrac{\omega}{\omega^{2}-3\omega+1-i(\omega^{2}-1)}.
\end{align} 
Eq.$(4.34)$ agrees with $x_{-}$ in Eq.(A$13)$ of Ref.$[1]$.
\item $\beta=-i\alpha$ case.\\
\begin{align}
\theta_{s}^{2}=\dfrac{\omega}{\omega^{2}-3\omega+1+i(\omega^{2}-1)}.
\end{align} 
Eq.$(4.35)$ also agrees with $x_{+}$ in Eq.(A$13)$ of Ref.$[1]$.
\end{enumerate}
Eqs.$(4.28)$, $(4.31)$, $(4.32)$, $(4.34)$, and $(4.35)$ complete the proof.\\
\par\noindent
{\bf Acknowledgments.}
NK acknowledges financial support of the Grant-in-Aid for Scientific
Research (C) of Japan Society for the Promotion of Science (Grant No. $21540116$).
\begin{small}
\bibliographystyle{jplain}

\end{small}
\noindent {\large{\bf Appendix A}}  \\
In Appendix, we provide with the proof of Lemma $1$.
From Eq.$(4.13)$, we have 
\begin{align}
\lambda\sum^{\infty}_{x=2}\Psi^{L}(x)z^{x}=\frac{1}{\sqrt{2}}\sum^{\infty}_{x=2}\Psi^{L}(x+1)z^{x}+\frac{1}{\sqrt{2}}\sum^{\infty}_{x=2}\Psi^{R}(x+1)z^{x},
\end{align}
and
\begin{align}
\lambda\sum^{\infty}_{x=2}\Psi^{R}(x)z^{x}=\frac{1}{\sqrt{2}}\sum^{\infty}_{x=2}\Psi^{L}(x-1)z^{x}-\frac{1}{\sqrt{2}}\sum^{\infty}_{x=2}\Psi^{R}(x-1)z^{x}.
\end{align}
Rewriting Eqs.$(4.36)$ and $(4.37)$ gives
\begin{align}
\left(\lambda-\dfrac{1}{\sqrt{2}z}\right)f^{L}_{+}(z)-\dfrac{1}{\sqrt{2}z}f^{R}_{+}(z)=\lambda z\Psi^{L}(1)-\dfrac{z}{\sqrt{2}}(\Psi^{L}(2)+\Psi^{R}(2))-\dfrac{1}{\sqrt{2}}(\Psi^{L}(1)+\Psi^{R}(1)),
\end{align}
and
\begin{align}
\left(\lambda+\dfrac{z}{\sqrt{2}}\right)f^{R}_{+}(z)-\dfrac{z}{\sqrt{2}}f^{L}_{+}(z)=\lambda\Psi^{R}(1)z.
\end{align}
From now on, we will rewrite Eqs.$(4.38)$ and $(4.39)$ in terms of $\Psi^{L}(0)$ and $\Psi^{R}(0)$ for the simplicity.
Eq.$(4.14)$ implies
\begin{align}
\Psi^{L}(2)+\Psi^{R}(2)=\sqrt{2}\lambda\Psi^{L}(1).
\end{align}
Hence RHS of Eq.$(4.38)$ is equivalent to 
\begin{align}
-\dfrac{1}{\sqrt{2}}\Psi^{L}(1)-\dfrac{\omega}{2\lambda}\Psi^{L}(0)+\dfrac{\omega}{2\lambda}\Psi^{R}(0).
\end{align}
Eq.$(4.14)$ also implies
\begin{align}
\Psi^{L}(1)=\sqrt{2}\lambda\Psi^{L}(0)-\dfrac{\omega}{\sqrt{2}\lambda}\Psi^{L}(0)+\dfrac{\omega}{\sqrt{2}\lambda}\Psi^{R}(0).
\end{align}
Combining Eq.$(4.38)$ with Eqs.$(4.40)$, $(4.41)$ and $(4.42)$, we see that RHS of Eq.$(4.38)$ is equivalent to 
$-\lambda\Psi^{L}(0)$.
Eq.$(4.14)$ also yields 
\begin{align}
\Psi^{R}(1)=\dfrac{\omega}{\sqrt{2}\lambda}\Psi^{L}(0)-\dfrac{\omega}{\sqrt{2}\lambda}\Psi^{R}(0).
\end{align}
Combining Eq.$(4.39)$ with Eq.$(4.43)$, we see that RHS of Eq.$(4.39)$ is equal to $\dfrac{\omega z}{\sqrt{2}}(\Psi^{L}(0)-\Psi^{R}(0))$.\\
Next, Eq.$(4.13)$ also implies
\begin{align}\lambda\sum^{-\infty}_{x=-2}\Psi^{L}(x)z^{x}=\frac{1}{\sqrt{2}}\sum^{-\infty}_{x=-2}\Psi^{L}(x+1)z^{x}+\frac{1}{\sqrt{2}}\sum^{-\infty}_{x=-2}\Psi^{R}(x+1)z^{x},\end{align}
and
\begin{align}\lambda\sum^{-\infty}_{x=-2}\Psi^{R}(x)z^{x}=\frac{1}{\sqrt{2}}\sum^{-\infty}_{x=-2}\Psi^{L}(x-1)z^{x}-\frac{1}{\sqrt{2}}\sum^{-\infty}_{x=-2}\Psi^{R}(x-1)z^{x}.\end{align}
For the same discussion as the positive parts, Eq.$(4.44)$ gives
\begin{align}
\left(\lambda-\dfrac{1}{\sqrt{2}z}\right)f^{L}_{-}(z)-\dfrac{1}{\sqrt{2}z}f^{R}_{-}(z)=\lambda\Psi^{L}(-1)z^{-1},
\end{align}
and Eq.$(4.45)$ yields
\begin{align}
\left(\lambda+\dfrac{z}{\sqrt{2}}\right)f^{R}_{-}(z)-\dfrac{z}{\sqrt{2}}f^{L}_{-}(z)=\lambda\Psi^{R}(-1)z^{-1}-\dfrac{z^{-1}}{\sqrt{2}}(\Psi^{L}(-2)-\Psi^{R}(-2))-\dfrac{1}{\sqrt{2}}\Psi^{L}(-1)+\dfrac{1}{\sqrt{2}}\Psi^{R}(-1).
\end{align}
Putting $x=0$ in Eq.$(4.13)$ gives
\begin{align}
\lambda\Psi^{R}(0)=\dfrac{1}{\sqrt{2}}\Psi^{L}(-1)-\dfrac{1}{\sqrt{2}}\Psi^{R}(-1),
\end{align}
and Eq.$(4.15)$ implies
\begin{align}
\Psi^{L}(-1)=\dfrac{\omega}{\sqrt{2}\lambda}\Psi^{L}(0)+\dfrac{\omega}{\sqrt{2}\lambda}\Psi^{R}(0).
\end{align}
Now combining Eq.$(4.39)$ with Eqs.$(4.15)$, $(4.48)$ and $(4.49)$, we obtain
\begin{align*}
\Psi^{R}(-1)=\dfrac{\omega}{\sqrt{2}\lambda}\Psi^{L}(0)+\dfrac{\omega}{\sqrt{2}\lambda}\Psi^{R}(0)-2\lambda\Psi^{R}(0).
\end{align*}
Consequently, RHS of Eq.$(4.44)$ is equal to $\dfrac{\omega}{\sqrt{2}z}(\Psi^{L}(0)+\Psi^{R}(0))$, and 
RHS of Eq.$(4.45)$ is identical to $-\lambda\Psi^{R}(0)$.
The discussions for positive and negative parts imply Lemma $1$.

\end{document}